\documentclass{aastex631}

\usepackage{CJK}
\usepackage{booktabs}  
\usepackage{multirow}  
\usepackage{makecell}
\usepackage{rotating}
\usepackage{tabularx}       
\usepackage{array}          
\usepackage{ragged2e}       
\usepackage{threeparttable}

\shorttitle{Formation of WASP-47}
\shortauthors{Wang et al.}

\begin{document}
\begin{CJK*}{UTF8}{gbsn}

\title{Formation of Ultra-short-period Planet in Hot Jupiter Systems: Application to WASP-47}

\correspondingauthor{Su Wang}
\email{wangsu@pmo.ac.cn}

\author[0000-0002-4859-259X]{Su Wang}
\affiliation{CAS Key Laboratory of Planetary Sciences, Purple Mountain Observatory,\\ Chinese Academy of Sciences, Nanjing 210008, China}
\affiliation{School of Astronomy and Space Science, University of Science and Technology of China, Hefei 230026, China}
\affiliation{CAS Center for Excellence in Comparative Planetology, Hefei 230026, China}

\author{Mengrui Pan}
\affiliation{Institute for Astronomy, School of Physics, Zhejiang University, Hangzhou 310027, China}
\affiliation{Center for Cosmology and Computational Astrophysics, Institute for Advanced Study in Physics, Zhejiang University, Hangzhou 310027, China}              
  
\author{Yao Dong}
\affiliation{CAS Key Laboratory of Planetary Sciences, Purple Mountain Observatory,\\ Chinese Academy of Sciences, Nanjing 210008, China}
\affiliation{CAS Center for Excellence in Comparative Planetology, Hefei 230026, China}

\author{Gang Zhao}
\affiliation{Nanjing Institute of Astronomical Optics \& Technology, Chinese Academy of Sciences, Nanjing 210042, China}
\affiliation{CAS Key Laboratory of Astronomical Optics \& Technology, Nanjing Institute of Astronomical Optics \& Technology, Nanjing 210042,China}

\author{Jianghui Ji}
\affiliation{CAS Key Laboratory of Planetary Sciences, Purple Mountain Observatory,\\ Chinese Academy of Sciences, Nanjing 210008, China}
\affiliation{School of Astronomy and Space Science, University of Science and Technology of China, Hefei 230026, China}
\affiliation{CAS Center for Excellence in Comparative Planetology, Hefei 230026, China}

\begin{abstract}
The WASP-47 system is notable as the first known system hosting both inner and outer low-mass planetary companions around a hot Jupiter, with an ultra-short-period (USP) planet as the innermost planetary companion. The formation of such an unique configuration poses challenges to the lonely hot Jupiter formation model. Hot Jupiters in multiple planetary systems may have a similar formation process with warm Jupiter systems, which are more commonly found with companions. This implies that the WASP-47 system could bridge our understanding of both hot and warm Jupiter formation. In this work, we propose a possible formation scenario for the WASP-47 system based on its orbital configuration. The mean motion resonance trapping, giant planet perturbations, and tidal effects caused by the central star are key factors in the formation of USP planets in multiple planetary systems with hot Jupiters. Whether a planet can become an USP planet or a short period super-earth (SPSE) planet depends on the competition between eccentricity excitation by nearby giant planet perturbations and the eccentricity damping due to tidal effects. The $Q_p'$ value of the innermost planet is essential for the final planetary configuration. Our results suggest that  a $Q_p'$ in the range of [1, 10] is favorable for the formation of the WASP-47 system. Based on the formation scenario, we estimate an occurrence rate of 8.4$\pm$2.4\%  for USP planets in systems similar to WASP-47. 
\end{abstract}


\section{Introduction} \label{sec:intro}
Among the thousands of confirmed exoplanets, a significant proportion consists of hot Jupiters (HJs). However, few HJs are detected to have planetary companions with orbital periods less than 50 days and radii smaller than 2 $R_\oplus$ \citep{Huang16}. Although at least six confirmed planetary systems, WASP-47, Kepler-730, WASP-132, TOI-1130, TOI-2000, and WASP-84 \citep{Becker15, 2018RNAAS...2..160Z, 2020ApJ...892L...7H, 2022AJ....164...13H, 2023MNRAS.524.1113S, 2023MNRAS.525L..43M}, as well as two candidate systems, TOI-5143 and TOI-2494 \citep{2023MNRAS.524.1113S} have been identified with both hot giant planets and nearby low-mass planetary companions, the occurrence rate of such planetary companions around HJs remains remarkably low compared to the total number of confirmed HJs. In contrast, more than half of the warm Jupiters (WJs) with orbital periods ranging from 10 to 200 days are accompanied by nearby planetary companions \citep{Huang16, 2023AJ....165..171W}. WASP-47, the first multiple planetary system detected with both a HJ and nearby planetary companions, represents a breakthrough in understanding the loneliness of HJs. This rare architecture offers us an opportunity to investigate the dynamical and formation processes of such systems. 

The WASP-47 planetary system contains a G9 star. The first discovered planet in the system is a hot Jupiter with an orbital period of 4.16 days (WASP-47b) \citep{Hellier12}. Subsequent K2 observations revealed a transiting super-Earth with an ultra-short period (USP) planet interior to the hot Jupiter (WASP-47e) and a Neptune-size planet with a longer orbital period outside the hot Jupiter (WASP-47d) \citep{Becker15}. Additionally, a long-period giant planet (WASP-47c) was monitored through radial velocity follow-up detection \citep{NV16}. The detailed parameters of the planets in the system are listed in Table \ref{system}. The configuration of WASP-47 can be seen as an extension of WJs into the HJs region, linking the formation processes of HJs and WJs \citep{Huang16}. The metallicity of the central star is (0.36 $\pm$ 0.05) dex, which is consistent with the observed higher occurrence rate of HJs around metal-rich stars \citep{2005ApJ...622.1102F, 2020MNRAS.491.4481O}.

Hot Jupiters are believed to be lonely without nearby planetary companions, suggesting a violent formation history. High-eccentricity migration is the most likely mechanism responsible for producing a lonely Jupiter on a short period orbit \citep{2018ARA&A..56..175D}, as this process can clear out material in the inner region of the system. However, the detection of inner planetary companions of HJs is another story from the lonely HJs. Observational data indicate that the three inner planets in the WASP-47 system are likely in a coplanar configuration \citep{Becker15, 2017MNRAS.468..549B, 2015ApJ...812L..11S}, suggesting a relatively quiet dynamical evolution process, possibly through in-situ or smooth orbital migration process. Nonetheless, the presence of the HJ in the WASP-47 system implies a substantial accumulation of solid material, which supports the orbital migration of a Jupiter-size planet from the outer region of the system. Additionally, the orbital period ratio between WASP-47 b and d, which lies within 20\% of a 2:1 mean motion resonance (MMR) \citep{Becker15}, further implies a consequence of an orbital migration process \citep{2002ApJ...567..596L, 2012ApJ...753..170W}. Therefore, there is a high likelihood that the planets in this system formed through an orbital migration process.

Another intriguing feature of the WASP-47 system is the presence of an USP planet in the system. 
The innermost planet, WASP-47 e, is a small and dense USP planet with a radius of 1.87 $R_{\oplus}$, indicating a possible rocky composition \citep{2017AJ....153..265W, 2016A&A...595L...5A}. Planets with radii around $1.6 R_{\oplus}$ are thought to be the critical boundary between smaller rocky planets and larger planets with substantial low-density atmospheres \citep{2015ApJ...801...41R}. However, these thresholds are based on planets that are not subjected to strong irradiation, like WASP-47 e, which is very close to the central star \citep{2015ApJ...801...41R, 2015ApJ...813L...9D}. Therefore, as an USP, WASP-47 e could either be a totally rocky composition planet or possess a rocky core surrounded by dense volatile materials, such as water, similar to 55 Cnc e \citep{2017AJ....154..237V}. This is another clue suggesting that WASP-47 e may have originated from a distant location and migrated to its current observed position. Most planetary systems with USP planets show large period ratios between the USP planet and its nearest neighbor \citep{2015ApJ...813L...9D}, with the remaining planets typically arranged in more compact configurations \citep{2018NewAR..83...37W}. This suggests that USP planets may have formed slightly farther out, closer to their neighboring planetary companions, before being affected by secular excitation or MMR trapping, which stirred up the eccentricity of the innermost planet \citep{2019AJ....157..180P}. Due to tidal interactions with the central star, the orbit of the innermost planet would be circularized and shrink to an orbit of less than one day, forming an USP planet. Therefore, the perturbation from the nearby gas giant, MMR trapping during orbital migration, and the tidal effect caused by the central star are three key processes determining whether a planet can become an USP planet.

\begin{figure*}
\centering
\includegraphics[scale=0.52]{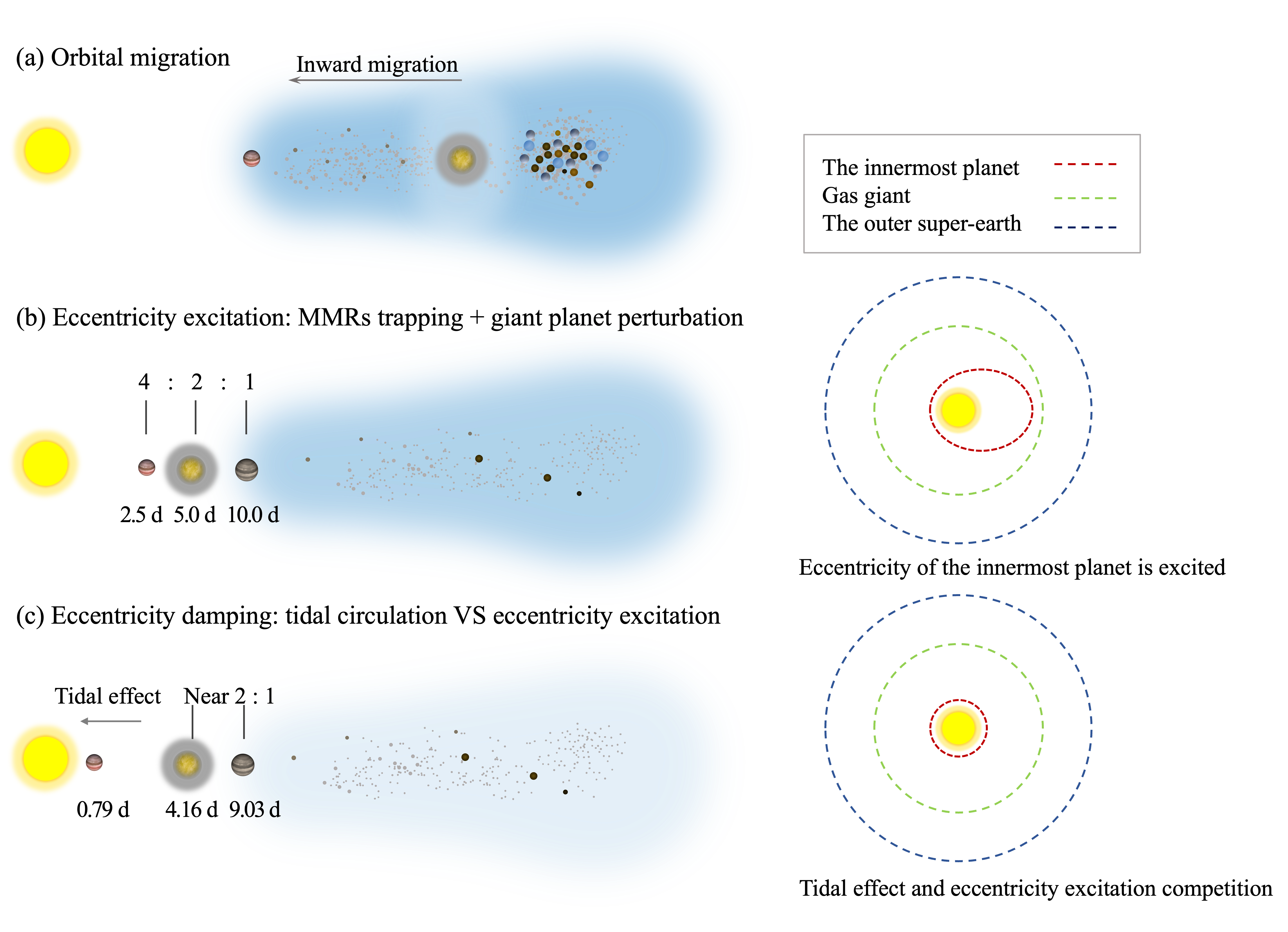}
 \caption{A schematic figure illustrates the main formation process of the WASP-47 system. Panel (a) shows the orbital migration process, where the innermost planet halts near the inner edge of the gas disk. Panel (b) depicts the eccentricity excitation due to MMRs trapping and giant planet perturbation. Panel (c) displays the eccentricity damping caused by the tidal effects and the semi-major axis of the innermost planet decreases gradually, ultimately forming an USP planet. The right panel presents the orbital configurations of the planetary systems at different formation stages. 
 \label{schematic}}
\end{figure*}

In this work, we propose a scenario to explain the formation of systems like WASP-47, which consists of a hot Jupiter with low-mass USP planetary companion. Figure \ref{schematic} provides a schematic illustration of the main formation process.
Firstly, planets form in the outer region of the protoplanetary disk, accumulating material and growing to nearly their current masses. Due to interactions with the gas disk, the planets begin to undergo orbital migration. The innermost planet experiences rapid type I migration, eventually halting near the inner edge of the gas disk. Subsequently, the giant planet and the outer super-Earth migrate inward, approaching the innermost planet. By the end of the migration process, the three planets become trapped in MMRs, with orbital periods of about 2.5, 5.0, and 10.0 days, respectively, leading to eccentricity excitation. Then the eccentricity of the innermost planet alternates between excitation and suppression because of the combined tidal effects raised by the central star and the gravitational influence of the giant planet, along with a gradual decrease of its semi-major axis. This process ultimately causes the innermost planet to migrate further inward, forming an USP planet (moving from 2.5 days to about 0.79 days), leaving the other two planets in a near 2:1 MMR configuration. The fourth planet, WASP-47 c, is distant from the inner three planets, and despite its eccentric orbit, the inner three planets are relatively unaffected by its perturbations \citep{2020AJ....159..176K}. Therefore, the influence induced by the outermost planet is ignored here (an estimation of its potential effect is discussed in Section \ref{sec:fourth}).

The paper is organized as follows: Section 2 describes the models, including the dynamical effects and numerical methods we used in this work, the main results are presented in Section 3, and the conclusions are summarized in Section 4.

\begin{figure*}
\centering
\includegraphics[scale=0.32]{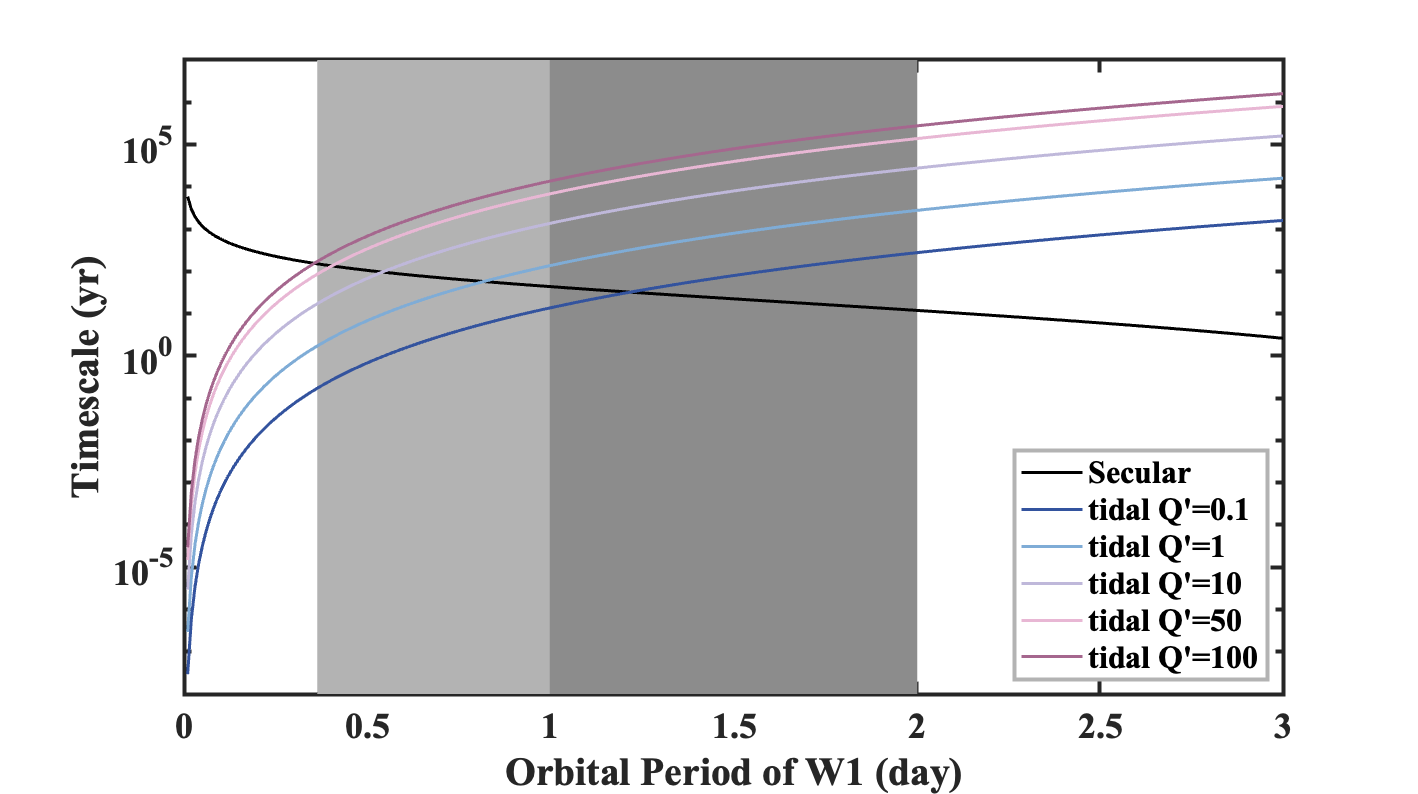}
 \caption{The timescales of tidal damping and the secular evolution. The light gray region represents the orbital period range $p\in [0.365, 1]$ day, while the dark gray region corresponds to $p\in [1, 2]$ day. 
 \label{dampt}}
\end{figure*}

\begin{table*}
\centering \caption{The parameters of the planets in the WASP-47 system (A: \citet{Becker15}, B: \citet{2017AJ....153...70S}, C: \citet{2017AJ....153..265W}).
 \label{tb1}}
\begin{tabular*}{18cm}{@{\extracolsep{\fill}}llllllll}
\hline
 Planet& Semi-major Axis& Orbital Period&Eccentricity&Radius&Mass&Density&Name\\
 &(AU)&(Day) && ($R_\oplus$)&($M_\oplus$)&$\rm gcm^{-3}$&\\
\hline
$W_1$&0.017 (A) &0.79 (A) &0.03 (C)&1.87 (B)&9.1 (C)&7.63 (B)&e\\
$W_2$&0.050 (A) &4.16 (A) &0.0028 (C) & 13.11 (B)&358 (C)&0.87 (B)&b\\
$W_3$&0.085 (A) &9.03 (A) &0.007 (C) & 3.71 (B)&13.6 (C)&1.36 (B)&d\\
$W_4$&1.382 (B) &595.7 (B) &0.28 (C) & &416 (C)&&c\\
\hline
\label{system}
\end{tabular*}
\end{table*}

\section{Models}
In this section, we study the main models considered in the formation process of the WASP-47 system. We estimate the dynamical effects in the model in Section \ref{sec:dynamical}, and the numerical methods and initial conditions are presented in Section \ref{sec:condition}. Herein, the planets in the system are designated as $W_1$, $W_2$, $W_3$, and $W_4$ in order of increasing distance from the central star. 
\subsection{Dynamical effects in the model}
\label{sec:dynamical}
\subsubsection{Effect of the gas disk}
\label{sec:gas}
We assume that the innermost planet ($W_1$) has already migrated close to the inner edge of the gas disk. Based on the estimation of the truncation radius of the gas disk, the inner edge is approximately nine times the stellar radius \citep{1991ApJ...370L..39K}. Given that the stellar radius is 2-3 times larger before reaching the main-sequence stage, the inner edge of the gas disk is likely to correspond to an orbital period of roughly 10 days or less. In this work, we set the inner edge at 8 days. The hot Jupiter ($W_2$) and the third planet  ($W_3$), still surrounded by the gas disk, are migrating toward the inner region of the system. Observations of young star clusters suggest the lifetime of the gas disk $\tau_{disk}$ is around a few million years \citep{Haisch01, WC17}. Herein, we assume the gas disk could persist for an additional 0.01-1 million years, taking into account that the protoplanetary disk has already evolved for the first few million years. The forces driving orbital migration $\textbf{F}_{\rm mig}$ and eccentricity damping $\textbf{F}_{\rm edamp}$ are expressed as \citep{CN06}
\begin{eqnarray}
\textbf{F}_{\rm mig}=-\frac{\textbf{v}}{\tau_{\rm a}},~~~~~~
\\
\textbf{F}_{\rm edamp}=-\frac{(\textbf{v}\cdot \textbf{r})\textbf{r}}{r^2 \tau_{\rm edamp}},
\end{eqnarray}
where $\textbf{v}$, $\textbf{r}$, and $r$ represent the velocity and position vectors of the planet, and the radial distance from the central star, respectively. $\tau_{\rm edamp}$ and $\tau_{\rm a}$ denote the eccentricity damping timescale and the orbital migration timescale, respectively. As the gas disk dissipates, its effect on the planets will be weakened by a factor of $e^{-t/\tau_{\rm disk}}$.

\subsubsection{Tidal effect acting on the planet}
\label{sec:star}
The tidal effect raised by the central star cannot be ignored due to the short orbital periods of the three planets in the system. The acceleration resulting from the tidal interactions can be expressed as \citep{ML02}
\begin{eqnarray}
\textbf{F}_{\rm tidal}=-\left(\frac{9n_p}{2Q'_p}\right)\left(\frac{M_*}{m_p}\right)\left(\frac{S_p}{a_p}\right)^5\left(\frac{a_p}{r_p}\right)^8
\nonumber\\
\times[3(\hat{\textbf{v}}\cdot \dot{\textbf{r}})\hat{\textbf{v}}+(\hat{\textbf{r}}\times \dot{\textbf{r}}-r\mathbf{\Omega}_p)\times \hat{\textbf{v}}],~~
\end{eqnarray}
where $n_p$, $Q'_p$, $S_p$, $a_p$, $r_p$, and $\mathbf{\Omega}_p$ represent the mean motion, effective tidal dissipation factor, radius, semi-major axis, distance from the central star, and spin rate of the planet, respectively.
The timescale for orbital circularization caused by tidal effects is given by the following expression \citep{ML02, 2008IAUS..249..285Z}

\begin{equation}
\tau_{cir}=\frac{4Q'_p}{63n_p}\left(\frac{m_p}{M_*}\right)\left(\frac{a_p}{S_p}\right)^5.
\label{tidal}
\end{equation}
The orbital circularization timescale depends on the planet's mass, semi-major axis, radius, and the tidal dissipation factor $Q_p'$. 

For the planets in the WASP-47 system, the planetary radii and masses have been measured  accurately \citep{Hellier12, Becker15, 2016A&A...595L...5A, 2017AJ....154..237V, 2017AJ....153...70S, 2017AJ....153..265W, 2023A&A...673A..42N}, the orbital circularization timescale is primarily governed by the effective tidal dissipation factor $Q_p'$ which becomes a crucial parameter in the formation model. In this work, we choose the value of $Q_p'$ based on estimates for planets in the Solar system \citep{1999ssd..book.....M} and super-Earths observed by the Kepler mission \citep{2019NatAs...3..424M}. Considering that the density of planet e is even higher than that of Earth, we choose a $Q_p'$ for planet e within the range of [0.1, 100]. For planet b which is similar to the gas giants in the Solar system, we choose $Q_p'$ within [$10^4, 10^6$]. $Q_p'$ for planet d is chosen within [0.1, $10^5$] based on the estimation of the value for super-earths.

\subsubsection{Secular perturbation caused by $W_2$}
The timescale for the eccentricity excitation of a planet due to secular interactions between two planets can be expressed as \citep{1999ssd..book.....M, 2003ApJ...598.1290Z}
 
\begin{equation}
\tau_{secular}=\frac{2\pi}{g_1-g_2},
\end{equation}
where $g_1$ and $g_2$ are the two eigenfrequencies. The black line in figure \ref{dampt} represents $\tau_{secular}$, which arises from the interaction between $W_1$ and $W_2$, with $W_2$ located around 4 days. The $x$-axis represents the orbital period of $W_1$. The eccentricity of a planet, which initially follows a circular orbit, can be excited to an equilibrium value by an outer perturber, as described in \citep{2007MNRAS.382.1768M}

\begin{equation}
e_p^{(eq)}=\frac{(5/4)(a_p)/(a_c)e_c\varepsilon_c^{-2}}{|1-\sqrt{a_p/a_c}(m_p/m_c)\varepsilon_c^{-1}|},
\label{secular}
\end{equation}
where $\varepsilon_c=\sqrt{1-e_c^2}$, with subscripts $p$ and $c$ referring to the parameters of the planet and its companion, respectively. Based on the estimation of $e_p$, the eccentricity of an inner planet with an orbital period of approximately 2.5 days can be excited to a maximum of about 0.3, given the presence of an outer giant planet with an orbital period of 5 days and an eccentricity of $e=0.2$.

\subsubsection{Effect of giant planet $W_4$ in the outer region}
\label{sec:fourth}
The giant planet in the outer region is located at approximately 1.4 AU. The secular resonance caused by the perturbation between the two giant planets may influence the formation of the three inner planets, especially when the proper frequency of the innermost planet coincides with one of the eigenfrequencies \citep{1980Icar...41...76H}. The sweeping of secular resonances can excite the eccentricities of the planets \citep{2017ApJ...849...98Z, 2017ApJ...836..207Z}, which could be a possible reason leading to the formation of an USP planet. As the outermost giant planet migrates inward from 3.5 AU to 1.387 AU, the locations of the secular resonances sweep from the position close to the central star, which is less than $10^{-5}$ AU to around $10^{-4}$ AU. However, the secular resonances do not pass through the region where the innermost planet is located. Therefore, in this work, we neglect the gravitational influence of the outermost planet on the three inner planets.

\subsection{Numerical methods and initial conditions}
\label{sec:condition}
In this work, we primarily consider the gravitational interactions between each other, the orbital migration and eccentricity damping induced by the gas disk prior to its dissipation, and the tidal effects caused by the central star. The acceleration of a planet with mass $m_i$ is given by 

\begin{eqnarray}
\frac{d}{dt}\textbf{V}_i =
 -\frac{G(M_*+m_i
)}{{r_i}^2}\left(\frac{\textbf{r}_i}{r_i}\right) ~~~~~~~~~~~~~~~
\nonumber\\
+\sum _{j\neq i}^N
Gm_j \left[\frac{(\textbf{r}_j-\textbf{r}_i
)}{|\textbf{r}_j-\textbf{r}_i|^3}- \frac{\textbf{r}_j}{r_j^3}\right]
\nonumber\\
+\textbf{F}_{\rm edamp}+\textbf{F}_{\rm mig}+\textbf{F}_{\rm tidal},~~~~~
\label{allequ}
\end{eqnarray}
where $\textbf{V}_i$, $\textbf{r}_i$, $r_i$, and $M_*$ represent the velocity and position vectors of planet $i$, the radial distance from the central star, and the mass of the central star, respectively.  We integrate equation (\ref{allequ}) to investigate the dynamical evolution process of the three planets in the system using the Rebound Code \citep {RT15}. 

We carry out 510 runs with systems containing one gas giant and two low-mass companions. The initial conditions are listed in Table \ref{initial}. $W_1$ is assumed to have already migrated to the inner edge of the gas disk and halted at about 8 days. $W_2$ begins its migration from 80 days. According to the estimation of the type II migration timescale for gas giants \citep{2008ApJ...673..487I}, $\tau_a$ of $W_2$ is approximately on the order of $10^4$ to $10^5$ years, and $\tau_{\rm edamp}$ of $W_2$ is $\sim$ 0.1$\tau_a$. Therefore, we set $\tau_a$ for $W_2$ to $10^4$ years. The initial location and the migration timescale of $W_3$ are uncertain. The initial positions of $W_3$ are chosen to be at 160, 170, 180, and 190 days, respectively. The migration timescales of $W_3$ are set to $0.7,~0.8,~0.9,~1.0\times10^4$ years, respectively.  All planets are initially assumed to be in coplanar, nearly circular orbits. The argument of pericenter and the mean anomaly are randomly generated between $0^\circ$ and $360^\circ$.

\section {Numerical Simulation Results}
\subsection{Forming WASP-47 system configuration}
\begin{figure*}
\centering
\includegraphics[scale=0.3]{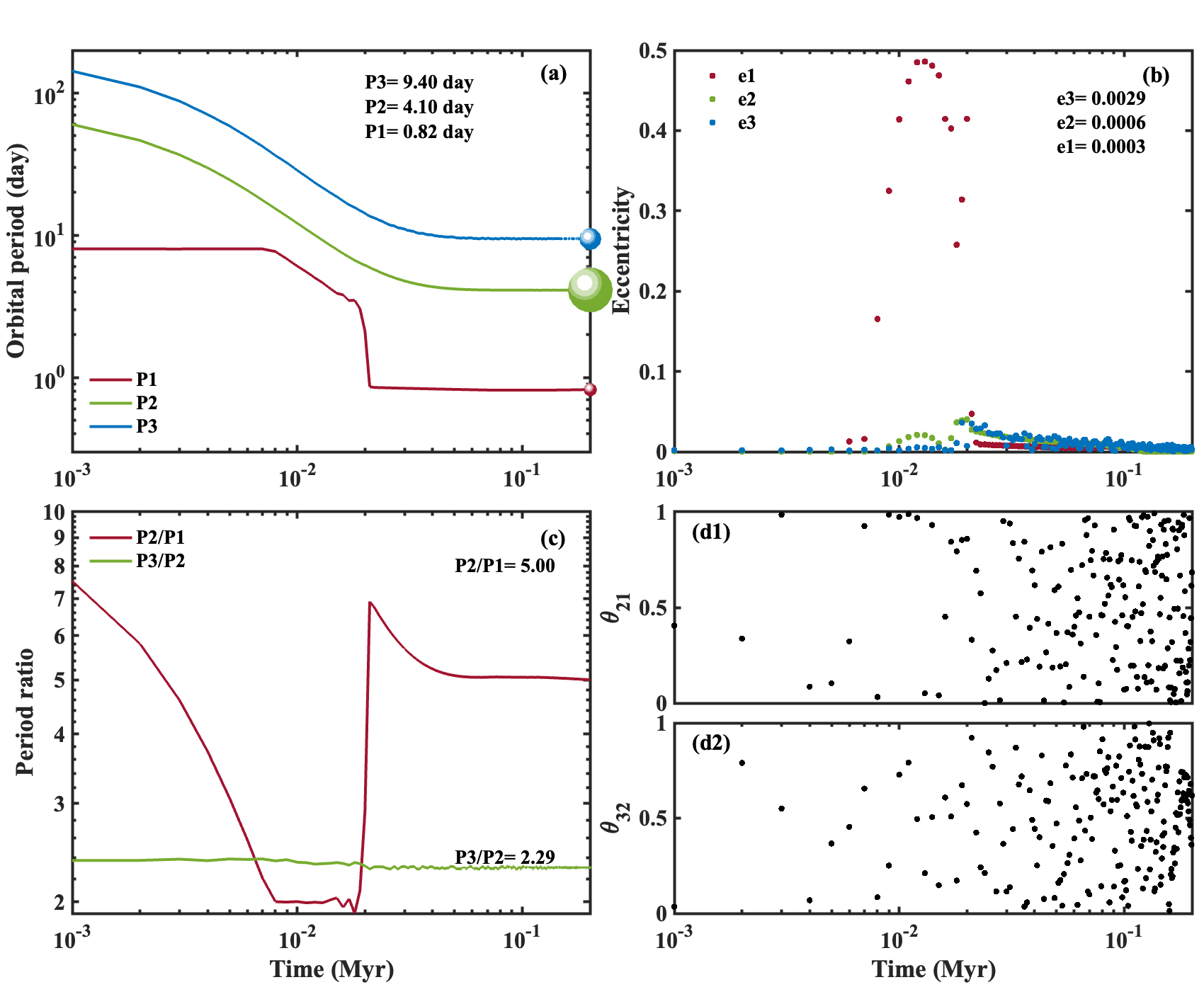}
 \caption{The orbital evolution of a typical run which forms a planetary configuration closely resembling the observed WASP-47 system. Panel (a) and (b) show the evolution of the orbital periods and eccentricities. The red, green, and blue lines and dots represent the evolution of $W_1$, $W_2$ and $W_3$, respectively. Panel (c) illustrates the evolution of the period ratios between adjacent planets. Panel (d1) and (d2) show the evolution of the resonant angles between neighboring planets, where $\theta_{21}=2\lambda_2-\lambda_1-\varpi_1$, and $\theta_{32}=2\lambda_3-\lambda_2-\varpi_2$.
 \label{typical}}
\end{figure*}

Based on the proposed formation scenario, we reproduced the formation process of the planetary configuration of the WASP-47 system through numerical simulations. Figure \ref{typical} shows a typical result from our simulations. In this case, $W_3$ starts at 190 days with a migration timescale of $10^4$ years. The $Q_p'$ for the three planets are set to 1, $10^6$, and $10^4$, respectively. As shown in figure \ref{typical}, $W_1$ is positioned at around 8 days, while $W_2$ and $W_3$ migrate inward from the outer region of the system. When $W_2$ approaches $W_1$, they become trapped into a 2:1 MMR at about $8\times 10^{-3}$ Myr. The orbital period ratio between $W_1$ and $W_2$ decreases from 7.5 to approximately 2.0. 

Since the period ratio of the outer two planets initially approaches 2.0, $W_2$ and $W_3$ become trapped into a near 2:1 MMR quickly. By the end of the orbital migration process, both planet pairs are trapped into 2:1 MMRs, with $W_3$ halting at around 9 days, which is near the edge of the protoplanetary disk. Panel (d1) shows the evolution of the resonant angles between $W_1$ and $W_2$. We can find that $\theta_{21}$ librates in the interval between $0.8\times10^{-3}$ and $2\times10^{-2}$ Myr, which is consistent with the evolution of the period ratio. 

With the MMRs trapping between the planets, their eccentricities are excited, especially that of $W_1$. Combined with the perturbation from the gas giant, the eccentricity of $W_1$ can be excited to nearly 0.5 at about 0.01 Myr. In this case, $Q'_p=1$ for $W_1$ leads to a very short timescale of tidal damping of the eccentricity. From figure \ref{dampt}, we can see that with $Q_p'=1$ (represented by the light blue line), the timescale for eccentricity excitation due to the perturbation of the outer giant shifts from being longer than the tidal damping timescale to comparable to it as $W_1$ migrates inward from 2.5 days. Thus, under the combined effects of the perturbation of the giant and eccentricity damping, the eccentricity of $W_1$ is excited more than once and eventually damped to near 0, making the innermost planet stabilized at a position of $\sim$ 0.8 days.

Since $W_2$ is the most massive of the three planets, its eccentricity is excited the least. Meanwhile, the eccentricity of $W_3$ can be excited to near 0.1. By the end of the simulation, the period ratio between $W_1$ and $W_2$ expands from near 2.0 to 5.0, while the period ratio between $W_2$ and $W_3$ remains close to 2.0, resulting in a final orbital configuration that closely resembles the observed WASP-47 system. The theoretically predicted positions of the three planets differ from observed values by less than 4\%.

\begin{table*}
\centering \caption{The initial parameters of the planets used in the simulations.
 \label{initial}}
\begin{tabular*}{18cm}{@{\extracolsep{\fill}}cccccccc}
\tableline
\tableline
 Planet& Orbital Period&Timescale of migration&$Q_p'$\\
&(Day) &(Year)& \\
\hline
1&8&-&0.1, 1, 10, 50, 100\\
2&80&$10^4$ & $10^4$, $10^5$, $10^6$\\
3&160, 170, 180, 190 &(0.7, 0.8, 0.9, 1.0)$\times 10^4$& 0.1, 1, 10, 100, $10^3$, $10^4$, $10^5$\\
\tableline
\tableline
\end{tabular*}
\end{table*}

\begin{figure*}
\centering
\includegraphics[scale=0.72]{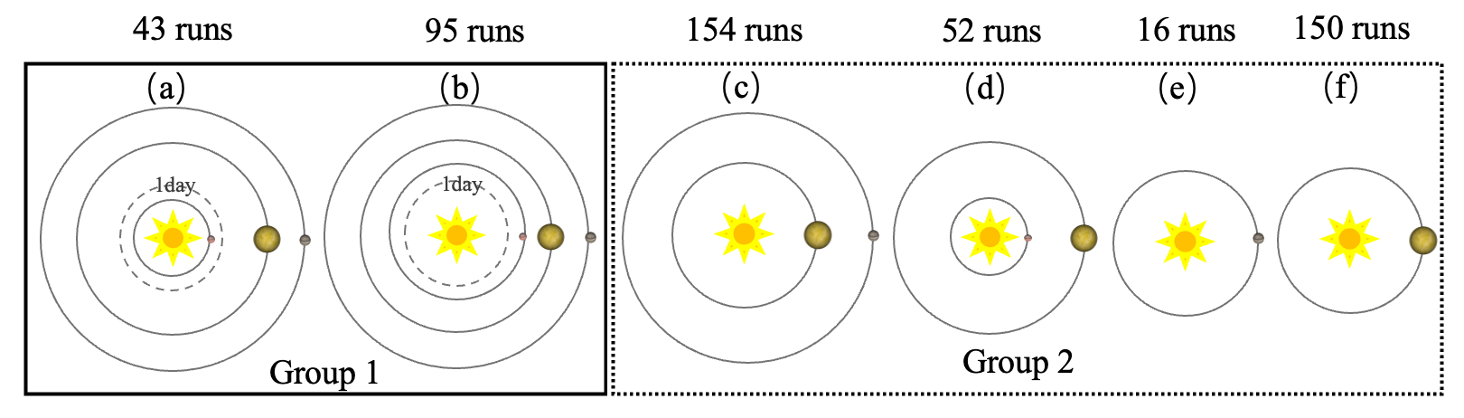}
 \caption{The final planetary configurations obtained from the numerical simulations.
 \label{config}}
\end{figure*}

\begin{figure*}
\centering
\includegraphics[scale=0.2]{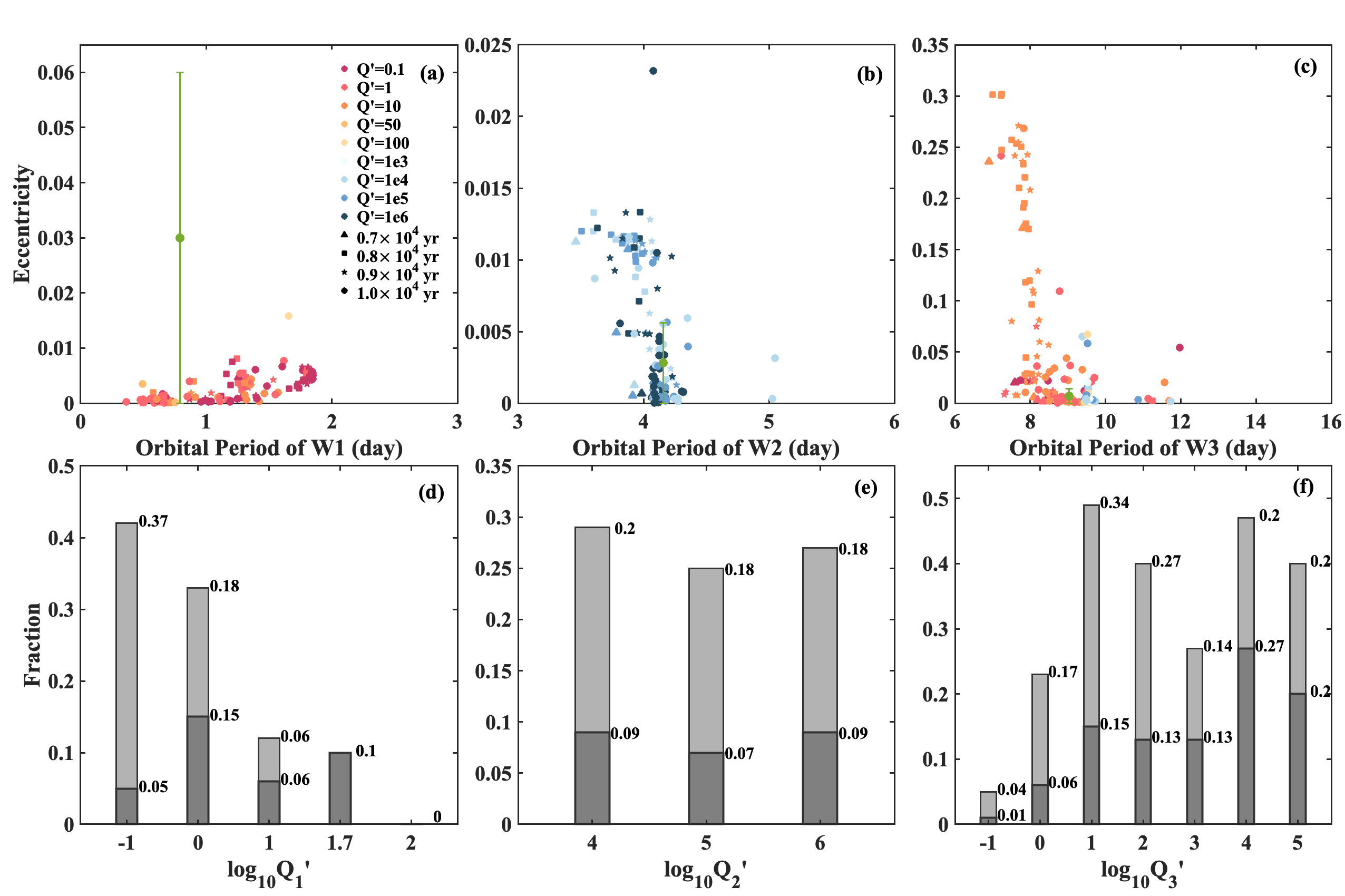}
 \caption{The final distributions of eccentricities and orbital periods for the three planets obtained from the simulations in group 1. Panel (a), (b), and (c) illustrate the orbital period distributions of $W_1$, $W_2$, and $W_3$, respectively. The colors represent the $Q_p'$ used for the planets, and the shapes of the dots indicate different migration timescales of the third planet, as shown in the legend in panel (a). The green dots and error bars represent the estimated values from observations. Panel (d), (e), and (f) show the number distributions of $Q_p'$ for the three planets. The light gray bars indicate the number distribution for runs in which SPSE planets are formed, while the dark gray bars represent the number distribution for the systems that formed USP planets. 
 \label{st1}}
\end{figure*}

\begin{figure*}
\centering
\includegraphics[scale=0.27]{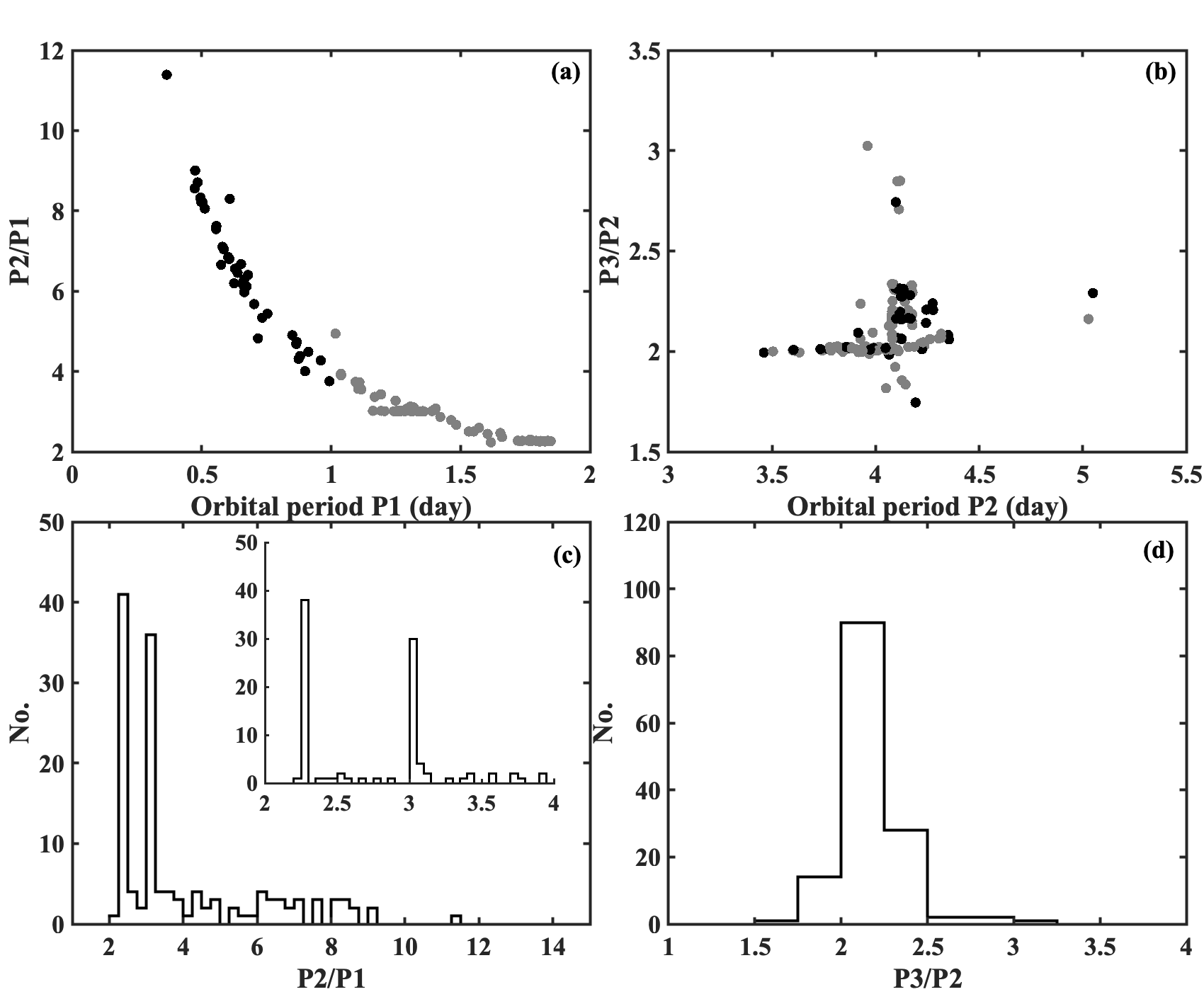}
 \caption{The distribution of orbital period ratios between adjacent planets in the system from group 1. Panel (a) shows the period ratio between $W_1$ and $W_2$ as it changes with the orbital period of $W_1$. Panel (b) displays the period ratio between $W_2$ and $W_3$ as it changes with the orbital period of $W_2$. Panel (c) and (d) present the number distribution of the period ratios. A zoomed-in view of the period ratio distribution between 2.0 and 4.0 is provided in panel (c). Black and gray dots represent systems in which an USP planet and a SPSE planet are formed, respectively.
  \label{st3}}
\end{figure*}

\subsection{Statistical results}
Based on the results from 510 runs, the final planetary configurations obtained from the simulations can be categorized into two groups. In group 1, three planets form a stable configuration, while in group 2, orbital crossing occurs between the planets. The possible planetary configurations are shown in figure \ref{config}. Although $W_1$ can even move to as close as 0.0015 AU (equivalent to 0.022 days), planets with semi-major axes less than 0.01 AU may be engulfed by the central star in the WASP-47 system. Therefore, we assume that any planet with a semi-major axis smaller than 0.01 AU (corresponding to an orbital period of 0.365 days) in a system like WASP-47 will eventually merge with the central star. According to our statistical results, USP planets with semi-major axes in the range of $a\in [0.012, 0.0195]$ AU form in 43 systems. The occurrence rate of USP planets in such configuration is about 8.4$\pm$2.4\%. The main results are summarized in table \ref{result}.

\begin{table*}
\centering 
\caption{The statistical results of 510 runs of numerical simulations. ``Config." means the planetary configurations formed as shown in figure \ref{config}. ``No. of planets" refers to the number of remaining planets in the systems. $P_i$ and $e_i$ are the orbital period and eccentricity of planet $W_i$ (i=1, 2, 3), respectively, while $P_{j+1}/P_j$ (j=1, 2) is the period ratio between two adjacent planets in the final configurations.
 \label{result}}
\begin{tabular}{c|c|c|c|c}
\tableline
\tableline
&&&&\\
& Config.&\parbox{1.0cm} {No. of runs}&\parbox{1.1cm} {No. of planets}& Main results\\
 &&&&\\
\hline
 \multirow{2}*{\parbox{3cm}{\centering Group1:\\ ~\\Stable configuration} } &\parbox{1.45cm}{Fig. \ref{config} (a)} & 43&3&{\parbox{9.3cm}{~\\$\bullet~$Form an USP planet in a three-planet system\\$P_1$:  [0.365-1.0] day ~~$e_1$: [$2\times 10^{-5}$-0.004]~~$P_2/P_1$: [3.8-11.4]\\$P_2$: $\sim$ 4.0 day ~~~~~~~~$e_2$: [0.0001-0.01]~~~~~~$P_3/P_2$: [1.7-2.7]\\
 $P_3$: [6.9-11.6] day ~~~$e_3$: [0.0008-0.25]\\~}}\\
        \cline{2-5}
           &     Fig. \ref{config}  (b)& 95 & 3&\parbox{9.3cm}{~\\$\bullet~$Form a SPSE in a three-planet system\\$P_1$: [1.0-1.85] day ~~~$e_1$: [0.0002-0.008]~~~~~$P_2/P_1$: [2.2-5.0]\\$P_2$: $\sim$ 4.0 day ~~~~~~~$e_2$: [$2.3\times 10^{-5}$-0.02]~~$P_3/P_2$: [1.8-3.0]\\
 $P_3$: [7.0-12.0] day ~~$e_3$: [$0.0005-0.30$]\\~} \\       
\hline
 \multirow{4}*{\parbox{3cm}{\centering Group2:\\~\\Orbital crossing happened} } &Fig. \ref{config} (c) & 154&2&\parbox{9.3cm}{~\\$\bullet~$One hot Jupiter with one outer companion configurations\\~}\\
        \cline{2-5}
           &      Fig. \ref{config} (d)& 52 & 2&\parbox{9.3cm}{~\\$\bullet~$One hot Jupiter with one inner companion configurations\\~} \\    
         \cline{2-5}
 \rule{0pt}{4ex}
           &      Fig. \ref{config} (e)& 16 & 1 &\parbox{9.3cm}{~\\$\bullet~$Only one short-period planet in the system\\
           $W_3$ survives\\
           an USP or SPSE planet\\~}\\  
                   \cline{2-5}
           &      Fig. \ref{config} (f)& 150 & 1 &\parbox{9.3cm}{~\\$\bullet~$A lonely hot Jupiter in the system\\
           11\% of $W_3$ scatter into outer space\\
           56\% of $W_1$ collide with the central star\\
           44\% of both low-mass companions merge with the Jupiter-size planet\\~}\\           
\tableline
\tableline
\end{tabular}
\end{table*}

\subsubsection{Group 1: three planets form in a stable configuration}

There are 138 systems in this group, with $W_2$ having an orbital period of around 4 days. The statistical results are shown in figure \ref{st1} - \ref{st3}. The eccentricities of $W_1$ are generally less than 0.008, with one exception where the eccentricity is approximately 0.015. The orbital periods of $W_1$ are all shorter than 1.85 days. For $W_2$, the eccentricities are mostly below 0.015, except for one case where the eccentricity reaches around 0.025. The orbital period of $W_2$ mainly concentrates around 4 days, with an error of $\sim$ 0.5 days. The eccentricity of $W_3$ can be excited to around 0.3, though in about 80\% of cases, the eccentricity of $W_3$ is less than 0.1. The period ratios between $W_1$ and $W_2$ fall within the range of [2.2, 11.4], as shown in panel (a) of figure \ref{st3}.

In group 1, 43 out of the 138 systems have $W_1$ with a period between 0.365 and 1.0 day, forming a so-called USP planet in the system (the configuration shown in panel (a) of figure \ref{config}), with $W_3$ forming at around 6.9-11.6 days. In 95 runs, the period of $W_1$ is larger than 1.0 day but less than 1.85 days (hereafter, we refer to these as short-period super-Earth planets, SPSE planets, with the configuration shown in panel (b) of figure \ref{config}). The final distribution of the planets formed in group 1 is shown in figure \ref{st1}. For the USP planets, eccentricities range between [$2\times10^{-5}$, 0.004], while the eccentricities of SPSE planets are in the range of [0.0002, 0.008], with one case reaching 0.015. 
The eccentricities of 66\% of $W_2$ are less than 0.006, which matches the range of eccentricity observed for the hot Jupiter in the WASP-47 system. And the eccentricities of 29\% of $W_2$ fall between 0.009 and 0.014. There is no significant difference in the eccentricities of hot Jupiters between systems with USP planets and SPSE planets, though in USP systems, the proportion of cases where $W_2$ has an eccentricity below 0.06 is slightly higher.
In approximately half of the 43 USP planetary systems, the eccentricities of $W_3$ exceed 0.014, which is the upper limit estimated from observational data for the third planet \citep{2017AJ....153..265W}. The eccentricities of $W_3$ can reach up to 0.25 in USP planetary systems, which is slightly lower than that in SPSE planetary systems. 

In group 1, the period ratio between $W_1$ and $W_2$ is greater than 2.2 and can extend up to 11.4, as shown in figure \ref{st3}, while the period ratio between $W_2$ and $W_3$ is mainly around 2.0, as seen in panel (d) of figure \ref{st3}. Panel (c) of figure \ref{st3} reveals two peaks in the distribution of the period ratio between $W_1$ and $W_2$, corresponding to the aggregation of SPSE planets with orbital periods near 1.3 and 1.7 days. The formation of USP planets is notably more challenging compared to SPSE planets in multi-planetary systems with a nearby giant companion.

Panel (d) to (f) of figure \ref{st1} show the distribution of $Q_p'$ for the planets in the systems of group 1. The systems with USP planets formed are shown in dark gray bars, while the distribution of $Q_p'$ in systems with SPSE planets formed is shown in light gray bars. For the formation of USP planets, the $Q_p'$ of $W_1$ is concentrated on the order of 1.0, whereas the formation of SPSE planets is more favorable with a $Q_p'$ of $W_1$ around 0.1, indicating a much quicker tidal circulation process. According to the estimation of tidal timescale shown in equation (\ref{tidal}) and the timescale of secular evolution shown in equation (\ref{secular}), we get the timescales vary with the semi-major axis, as shown in figure \ref{dampt}. In the region where $p\in [0.365, 1]$ days, the timescale of the secular evolution is comparable to the tidal circulation timescale with $Q_{p1}'=1$. This allows for the excitation of eccentricity, leading to the formation of USP planets. When $Q_{p1}'\ge10$ and $W_1$ migrates from 2 days to the inner region, the tidal circulation timescale exceeds the secular evolution timescale. Thus, in these cases with $Q_{p1}'\ge10$, the tidal effect is not efficient enough to suppress eccentricity excitation, resulting in more than 75\% of cases being unstable. In the remaining cases, most are unable to excite eccentricities significantly, leading to the formation of SPSE planets with smaller reductions in semi-major axis. The parameter $Q_{p2}'$ does not show a strong correlation with the formation of USP or SPSE planets. However, for systems similar to WASP-47, where the innermost planet forms at around 0.79 days, the $Q_p'$ of the hot Jupiters tends to be around $10^6$, similar to that of Jupiter in the Solar system. A larger $Q_{p3}'$ facilitates the formation of both USP and SPSE planets, especially when $Q_{p3}'$ exceeds 10. Therefore, the formation of USP planets is more dependent on the tidal process of the innermost planet.

From figure \ref{st1}, we find that the formation of short period planets is not significantly correlated with the initial position of $W_3$. USP planets and SPSE planets are more likely to form with a longer migration timescale of $W_3$, comparable to the migration timescale of $W_2$.

In group 1, the planetary configurations shown in panel (a) and (b) of figure \ref{config} can form. USP planets tend to form when $Q_p \in$ [1, 10], while SPSE planets are more likely to form with $Q_p \sim$ 0.1. Short-period planetary companions in multiple planetary systems are more likely to form when the eccentricity of a nearby giant planet is below 0.015. The orbital migration speed and the initial relative position of the outermost planet show no significant correlation with the formation of USP planets.

\subsubsection{Group 2: orbital crossing happened between the planets}
A total of 372 runs fall into group 2, where merging, collisions, or scattering events occurred, leaving behind a lonely Jupiter or a hot Jupiter with one planetary companion in the system. According to our statistical results, 150 runs resulted in only a surviving hot Jupiter (the planetary configuration is shown in panel (f) of figure \ref{config}), accounting for 29\% of all simulations. In these systems where a lonely hot Jupiter formed, 11\% of $W_3$ scattered into outer space, 56\% resulted in $W_1$ colliding with the central star, and 44\% led to the merger of both low-mass companions with the Jupiter-sized planet. 
The planetary configuration, which is similar to that shown in panel (d) of figure \ref{config}, can be formed if $W_3$ is scattered into outer space or collides with the hot Jupiter. This configuration leaves one short-period terrestrial planet and one hot Jupiter. It can be formed in 52 runs. Additionally, in group 2, 154 runs produced a system with one hot Jupiter and one outer companion, as illustrated in panel (c) of figure \ref{config}. Another possible outcome is a system with only one short-period planet, formed in 16 runs (as shown in panel (e) of figure \ref{config}). In these cases, the hot Jupiter and the innermost planet are engulfed by the central star, leaving $W_3$ as the only survivor, which could either be an USP or SPSE planet. Therefore, when orbital crossing happens between planets, the system is more likely to evolve into either a lonely hot Jupiter (29\%) or a hot Jupiter with one outer lower-mass planetary companion (30\%).

The eccentricity of $W_1$ can be damped to below 0.05 in stable cases, whereas in group 2, over half of the systems show eccentricities of $W_1$ being excited to larger than 0.1, with some cases reaching nearly 0.6, which is a key factor contributing to system instability. While most eccentricities of $W_2$ are below 0.1, 60\% of $W_2$ cases with the eccentricities exceeding 0.015, which is the upper limit of the eccentricity excited in group 1. As a result, the eccentricities of $W_2$ in group 2 are much greater than those in stable cases from group 1. Additionally, the eccentricity of $W_3$ in half of the cases in group 2 can be excited to greater than 0.3, leading to orbital overlap with the middle planet and triggering system instability.

\subsection{Comparison of numerical results and observations}
\begin{figure*}
\centering
\includegraphics[scale=0.24]{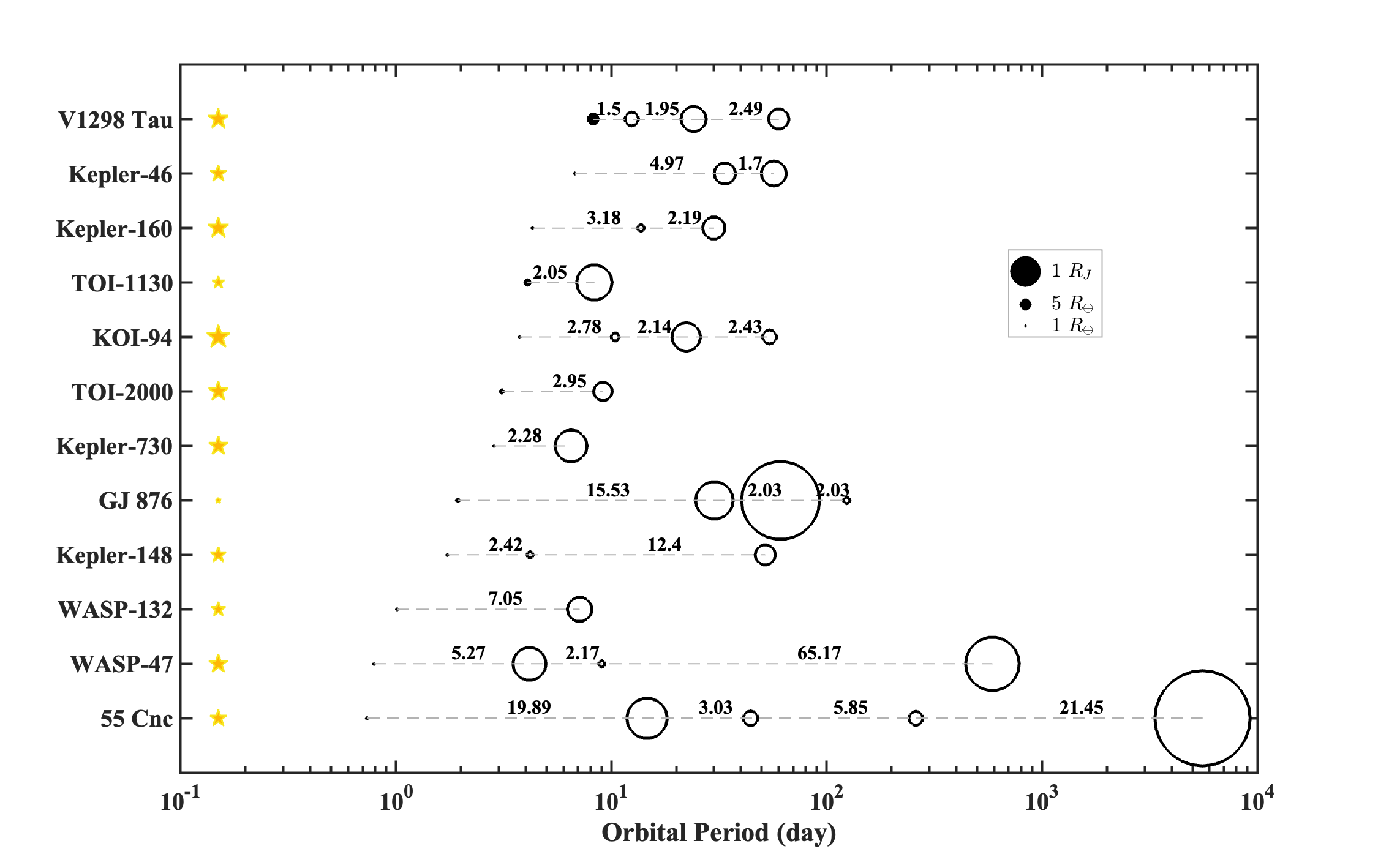}
 \caption{Confirmed planetary systems which have one giant planet accompanied by nearby low-mass planetary companions.
 \label{observation}}
\end{figure*}

Figure \ref{observation} presents confirmed planetary systems containing one giant planet and its nearby low-mass planetary companions. The orbital period of the low-mass planetary companion is less than 10 days, which is strongly influenced by the tidal effect from the central star. The period ratio between the innermost planet and its companion is far from resonance positions, suggesting a similar formation process to WASP-47. Observational data in figure \ref{observation} show that most innermost planets have orbital periods exceeding 1 day, consistent with our simulation results, which suggest that SPSE planets (19\%) are more likely to survive in such systems than USP planets. The strong tidal effect likely plays a significant role in shaping the inner SPSE planets of systems like Kepler-46 and WASP-132. 

\section{Conclusions and Discussions}

WASP-47 is the first planetary system observed to have both inner and outer low-mass planetary companions near a hot Jupiter \citep{Hellier12, Becker15, NV16}. This system is notable for its configuration, featuring two planets near MMRs and one USP planet. In this work, we propose a possible formation scenario for such systems, finding that MMR trapping, perturbations from the giant planet, and tidal effects are key factors in the formation of USP planets in multiple planetary systems.  

\begin{itemize}
\item
A planetary configuration similar to that of WASP-47 can form due to the combined effects of perturbations from the giant planet, mean motion resonance capture, and tidal effects caused by the central star. The eccentricity of the USP planet can be excited by the perturbations of the giant planet and MMR trapping, while it is damped by the tidal effect induced by the central star. The competition between eccentricity excitation and damping is crucial in determining the final position of the USP planet.

\item
USP planets tend to form within the range of [0.3, 1] days, while SPSE planets are more commonly found in the range of [1, 1.85] days, with perturbations from a giant planet around 2.5 days. Whether a short-period planet can evolve into an USP planet strongly depends on the tidal process between the planet and the central star. With an efficient tidal effect and a $Q_p'$-value less than 1, planets in multiple planetary systems with a nearby giant planet are more likely to evolve into SPSE planets. While with a $Q_p'$-value in the range of [1-10], short-period planets can evolve into USP planets, like WASP-47 e. 

\item
In this work, $Q_p'$ is expressed as $Q_p'=Q(1+19\mu/2g\rho S)$, where $\mu$ represents the planet's rigidity, $g$ is the gravitational acceleration at the surface of the planet, $\rho$ and $S$ are the mean density and radius of the planet, respectively \citep{ML02}. Estimates of Q-values for planets in the Solar system are not well-constrained and vary widely, typically ranging from [10, 200] \citep{1966Icar....5..375G, 1999ssd..book.....M}. Our estimates of $Q_p'$ for WASP-47 e suggest a value close to that of Earth, implying a similar layered structure, including core, mantle, and crust. Considering the planet's thermal state and response to tidal forces, the $Q_p'$-value is related to the internal properties of the planet, such as viscosity and rigidity \citep{2012A&A...541A.165R, 2015A&A...573A..23R}, especially when the interior is composed of layers with different materials \citep{2018A&A...613A..37B}. Combining our dynamical estimation on $Q_p'$ with possible compositions could provide a more detailed and precise picture of WASP-47 e's interior. 

\item
Under the formation scenario proposed in this work, five types of planetary configurations can form. We find that the possibility of forming a planetary system similar to WASP-47 is approximately 8.4$\pm$2.4\%. In most cases, the system evolves into a configuration with a lonely hot Jupiter, which is a commonly observed outcome in short-period exoplanet systems \citep{2018ARA&A..56..175D}. Another potential outcome is a hot Jupiter with an outer planetary companion, which could be identified in future observations.

\end{itemize}

Our simulation results suggest that the presence of a giant planet is a key driver of eccentricity excitation in the innermost planet. The mass, location, and orbital characteristics of the hot giant planet heavily influence the final system configurations. Additionally, the presence of low-mass planets between the USP planet and the giant planet may alter the final configuration of the system. We will explore these factors in greater details in future work.

\section*{Acknowledgments}
We thank the referee for a thorough and constructive report that significantly improved the manuscript. This work is supported by the National Natural Science Foundation of China (grant Nos. 12473076, 12033010, 12111530175, and 11873097), the Natural Science Foundation of Jiangsu Province (grant No. BK20221563), the B-type Strategic Priority Program of the Chinese Academy of Sciences (grant No. XDB41000000), the Foreign Expert Project (grant No. S20240145), Youth Innovation Promotion Association and the Foundation of Minor Planets of Purple Mountain Observatory.

\end{CJK*}
\end{document}